\documentclass[aps,prb,reprint,twocolumn,showpacs,floatfix]{revtex4-1}
\usepackage{graphicx}

\begin{document}

\title{Chiral spin liquid in a two-dimensional two-component helical magnet}

\author{Olga Dimitrova}

\affiliation{Institut f\"{u}r Theoretische Physik, Universit\"{a}t zu K\"{o}ln, Z\"{u}lpicher Str. 77, D-50937 K\"{o}ln, Germany}
\affiliation{Bah\c{c}e\c{s}ehir University, Y{\i}ld{\i}z Mh., 34349 Be\c{s}ikta\c{s}/\.{I}stanbul, Turkey}
\date{\today}

\begin{abstract}
A low-temperature method is developed, suited for the two-dimensional two-component classical helical magnet. Four phases on the phase diagram as functions of temperature and helicity parameter of the Hamiltonian are found. Among the three ordered phases two show magnetic order: the usual algebraic correlations of the magnetization and the algebraic correlations of the magnetization in the frame rotating according with the helical order. A chiral spin liquid phase emerges directly from the paramagnetic phase and has a scalar parity-breaking pitch of the magnetization as the order parameter. The chiral phase transition is found to be of a continuous second order type with a modified by the long-range interaction Ising universality class. All the critical exponents are calculated in the second and the third order of an $\epsilon$ expansion. A new scaling relationship replacing the Josephson's one is found. 
\end{abstract}


\maketitle

\section{Introduction}
\label{Introduction}

Magnetic systems with a broken parity and unbroken time-reversal symmetry, the chiral spin liquids, are predicted to possess excitations with unusual properties, like the anyon statistics~[\onlinecite{SpinLiquid}]. Spin liquids have been suggested to underpin the high-temperature superconductivity~[\onlinecite{SpinLiquid}] and they have been extensively searched for~[\onlinecite{Galitskii}]. On the other hand, the well known helical magnets demonstrate explicit parity violation 
in the ground state~[\onlinecite{HelicalMagnets, HelicalMagnets2, HelicalMagnets3, HelicalMagnets4}]. The usual path for a magnetic helix to form is via a single continuous phase transition~[\onlinecite{Kaplan}], where the chiral magnetic order develops. In two dimensions, Villain has proposed~[\onlinecite{Villain}] that the phase transition into a helix state can be split into two phase transitions, with the chiral, parity breaking phase transition taking place at a higher temperature, while the magnetic phase transition, breaking the time-reversal symmetry, taking place at a lower temperature. In between these two phase transitions a chiral spin liquid can exist~[\onlinecite{Villain}].

There is a simple model on a square lattice with nearest and frustrating next-nearest exchange interactions~[\onlinecite{Kaplan},\onlinecite{Doniach}], that has a helical ground state. The original study~[\onlinecite{Doniach}] and the follow-up study~[\onlinecite{Okwamoto}] of this simple model have indeed found the two phase transitions but in a usual sequence of events, i.e., they have found (incorrectly) that the helical order develops from the already robust magnetic order when lowering the temperature, i.e., the time-reversal phase transition takes place at a higher temperature than the parity-breaking phase transition. 

However, recent experiments~[\onlinecite{Cinti}] have found the reverse sequence of the two split phase transitions in a quasi-one-dimensional magnet, thus, validating the original Villain's picture of an intermediate chiral spin liquid. This reverse sequence was also proposed as a hypothesis, without a proof, in the context of the one-dimensional quantum spin chains~[\onlinecite{Kolezhuk}]. The same transition
sequence, i.e., chiral-BKT, is well established for a number of closely related models, e.g., fully frustrated XY model~[\onlinecite{Olsson}] and XY triangular AFM~[\onlinecite{Capriotti}]. Precisely the same model as considered in the present paper was recently studied by Sorokin {\it et al.}~[\onlinecite{Sorokin}] by classical Monte Carlo simulations. They find the same sequence of phase transitions as in the
present analytical study, however, the universality class of the chiral phase transition is established different. Although in the experiment~[\onlinecite{Cinti}] the magnetic phase transition has been determined to be of the Berezinskii-Kosterlitz-Thouless type~[\onlinecite{B,KT,K}], a few interesting questions about the chiral phase transition, like the critical exponents and the universality class, remained unanswered.  

The substantial uncertainty on the phase diagram of the simplest helical magnet calls for a revisit of the Garel and Doniach model~[\onlinecite{Doniach}]. In this paper, we develop an accurate low-temperature expansion method suited for this model. The Hamiltonian of the two-dimensional two-component frustrated magnet defined microscopically on the square lattice depends on one parameter called helicity. Applying our method for small helicity, in the long-range continuous approximation, we study thermodynamic phases and construct an accurate phase diagram in the plane of the temperature and the helicity parameter. One thermodynamic phase found is the chiral spin liquid. It has the scalar parity-breaking pitch of the magnetization, the vector otherwise random, as the order parameter.

We find that the chiral phase transition proceeds directly from the paramagnetic phase. In this situation, the magnetization vector is destroyed by strong vortex fluctuations and we find a representation of the model in terms of the pitch-field only. Using the Wilson's renormalization group and the $\epsilon$-expansion analysis, we find that the chiral phase transition in the Garel and Doniach model falls in the same universality class as the Ising model with long-range dipolar interactions~[\onlinecite{BrezinZinnJustin,Larkin,Aharony}]. We calculate the critical exponent $\eta$ up to the $\epsilon^3$ power of the $\epsilon$ expansion.

This paper is organized as follows. In Sec.~\ref{Model}, we introduce the model of the two-dimensional two-component, XY, helical magnet. In Sec.~\ref{Mean-field}, we derive the mean-field free energy density functional, which describes the helical magnet in a wide region of temperatures both below and above the chiral phase transition, except for the narrow critical region, where the fluctuations are highly developed. We prove that the long-range spin stiffness vanishes on the chiral phase transition. In Sec.~\ref{SectionPhaseDiagram}, we plot the phase diagram of the two-dimensional XY helical magnet as a function of the temperature and the helicity parameter of the Hamiltonian. On the phase diagram we find analytically four phases [shown in Fig.~\ref{phasediagram}]: two magnetically ordered phases, (i) with the two-dimensional XY algebraic correlations of the magnetization and (ii) with the algebraic correlations of the magnetization in the frame rotating according with the helical order, (iii) a 
usual disordered paramagnetic phase and emerging directly from it (iv) chiral spin liquid phase, which has the scalar parity-breaking chiral gradient of the magnetization, the so called helical pitch-field order, as the order parameter.
In Sec.~\ref{RG}, we go beyond the mean-field approximation and study the critical behavior at the chiral phase transition via a renormalization group analysis. We calculate the critical exponents in the second and the third order of a spatially anisotropic $\epsilon$ expansion. 

\section{Model}
\label{Model}

Garel and Doniach have introduced a model~[\onlinecite{Doniach}], consisting of classical XY spins: $\mathbf{S}_r=(\cos \phi_r,\sin\phi_r)$, 
at the sites $r$ of a square lattice interacting via the following exchange Hamiltonian:
\begin{equation}\label{Hamiltonian}
H=-\frac{1}{2}\sum_{r,r'} J_{rr'} \mathbf{S}_r\cdot \mathbf{S}_{r'},
\end{equation}
where the sum is restricted to the nearest neighbors in the $x$ and $y$ directions $(J_1)$ and to the next-nearest neighbors in the $x$ direction $(J_2)$. Furthermore, it is assumed that
$J_1>0$, $J_2< 0$ and $|J_2| > J_1/4$. The next-nearest neighbor antiferromagnetic interaction introduces a substantial frustration to the spin order provided $J_2\approx-J_1/4$. We study, in this paper, the properties of this model in the low-temperature region, where the spin directions change slowly on the lattice. In the continuum limit, the Hamiltonian can be cast as a functional of the smooth over the lattice magnetization function ${\mathbf m}(\mathbf{r})$:
\begin{eqnarray}\label{Hm}
{\cal H}=\frac{J}{2}\int d^2{\bf r}
\left[-\frac{\theta^2}{2}\left(\partial_x {\mathbf m} \right)^2+
\frac{a^2}{4}\left(\partial_x^2 {\mathbf m} \right)^2+\left(\partial_y {\mathbf m} \right)^2
\right],\nonumber\\
\end{eqnarray}
where ${\mathbf m}(\mathbf{r})=(m_x,m_y,0)$ is a unit vector ${\mathbf m}\cdot {\mathbf m}=1$, $a$ is the lattice constant and
\begin{eqnarray}\label{thetaJ}
\theta=\arccos{\frac{J_1}{4|J_2|}}
\end{eqnarray}
is the energetically preferred angle between the two adjacent spins along the $x$ direction, the so-called helix pitch. The continuum approach is valid provided the pitch angle $\theta\ll 1$. Via the replacement $m_x(\mathbf{r})+i m_y(\mathbf{r})=e^{i\phi(\mathbf{r})}$ 
the Hamiltonian~(\ref{Hm}) can be rewritten as:
\begin{equation}\label{Hphi}
{\cal H}=\frac{J}{2}\int \!\! d^2{\bf r}
\left\{
\frac{a^2}{4}\left[\left(\partial_x\phi\right)^2-q^2\right]^2\!\!+\!
\left(\partial_y\phi\right)^2+\!\frac{a^2}{4}\left(\partial_x^2\phi\right)^2
\right\},
\end{equation}
where $q=\theta/a$ is the pitch wave vector. At zero temperature the ground state of this Hamiltonian, with zero ground energy, has either of the two equivalent helical structures with $\phi(\mathbf{r})=\pm q x$: 
\begin{equation}
{\mathbf m}={\mathbf e}_x \cos{q x}\pm{\mathbf e}_y \sin{q x}.
\end{equation}
The ground state has a broken $Z_2$ symmetry. At an arbitrary temperature, the pitch-field order parameter of the helical state, the helix wave vector $Q({\bf r})$ is slowly varying in space. Casting the helical magnet model in terms of the new spin field:
\begin{equation}\label{Shift}
\psi({\bf r})=\phi({\bf r})-\int_{-\infty}^x Q({\bf r}) dx,
\end{equation}
we find the following Hamiltonian:
\begin{eqnarray}\label{HpsiQ}
{\cal H}&=&{\cal H}_0+{\cal H}_{int},\nonumber \\
\nonumber \\
{\cal H}_0&=&\frac{J}{2}\int d^2{\bf r}
\left[a^2\frac{3 Q^2-q^2}{2}\left(\partial_x\psi\right)^2\right.\nonumber \\
&&+\left(\partial_y\psi+\int_{-\infty}^x\!\! \partial_yQ\,\, dx\right)^2+\frac{a^2}{4}\left(\partial_x^2\psi+\partial_xQ\right)^2
\nonumber \\
&&+
\left.\frac{a^2}{4}(Q^2-q^2)^2+
a^2 Q (Q^2-q^2)\partial_x\psi\right],\nonumber \\
{\cal H}_{int}&=&\frac{J}{2}\int d^2{\bf r}
\left[
a^2 Q \left(\partial_x\psi\right)^3+\frac{a^2}{4}\left(\partial_x\psi\right)^4
\right],
\end{eqnarray}
suited for the low-temperature analysis. In the ground state, $Q(T=0)=q=\theta/a$. 
The bare Green's function at zero temperature is the propagator of the free, 
non-interacting spin-wave Hamiltonian ${\cal H}_0$: 
\begin{equation}\label{G0pQT0}
{\cal G}_0^{-1}(\mathbf{p})=\theta^2 p_x^2 +p_y^2 +\frac{a^2}{4}
p_x^4.
\end{equation}
The propagator of ${\cal H}_0$ at a finite temperature is
\begin{equation}\label{G0pQ}
G_0^{-1}(\mathbf{p})=\frac{3 Q^2 a^2-\theta^2}{2} p_x^2 +p_y^2 +\frac{a^2}{4}
p_x^4.
\end{equation}
At zero temperature, ${\cal G}_0(\mathbf{p})=G_0(\mathbf{p},T=0)$. There are two spin-wave interactions in the Hamiltonian~(\ref{HpsiQ}): the cubic and the quartic vertices. In the following sections, we use the normalized temperature and a unit lattice constant:
\begin{equation}
t=\frac{T}{J}, \quad a=1.
\end{equation}
In the low-temperature limit, many Fourier transformed integrals can be extended to the infinite momenta, however, some ultraviolet divergent integrals need to be taken inside the Brillouin zone of the square lattice: $p_x,p_y \in (-\pi,\pi)$. We assume the pitch field $Q({\bf r})$ to be uniform in the next section, whereas in the vicinity of the chiral phase transition it develops into a fluctuating inhomogeneous order parameter field.

\section{Low temperature expansion}
\label{Mean-field}

\begin{table}
\caption{Free energy diagrams.}
\centering
\begin{tabular*}{\linewidth}{p{0.08\textwidth} p{0.14\textwidth} p{0.14\textwidth} c}
\hline
\hline \\[-2.5ex]
& \quad\,\,\,\, 1 & \quad\, $Q^2$ & $Q^4$
  \\[0.5ex]
\hline 
\\[-2ex]
$T$& $\quad\frac{3}{8} \,(0,1)$ & $-\frac{3}{4}\, (2,0)$ & -- \\  [2ex]
$T^2$& $-\frac{3}{16} (0,2)$ & $+\frac{27}{8} (2,1)$ & $-\frac{27}{8}\,(4,0)$\\ [2ex]
$T^3$& $+\frac{9}{16} (0,3)$ & $-\frac{81}{16} (2,2)$ & $+\frac{243}{8}(4,1)$\\ [1.5ex]
\hline\hline
\end{tabular*}
\end{table}

\begin{figure} 
\includegraphics[width=0.45\textwidth]{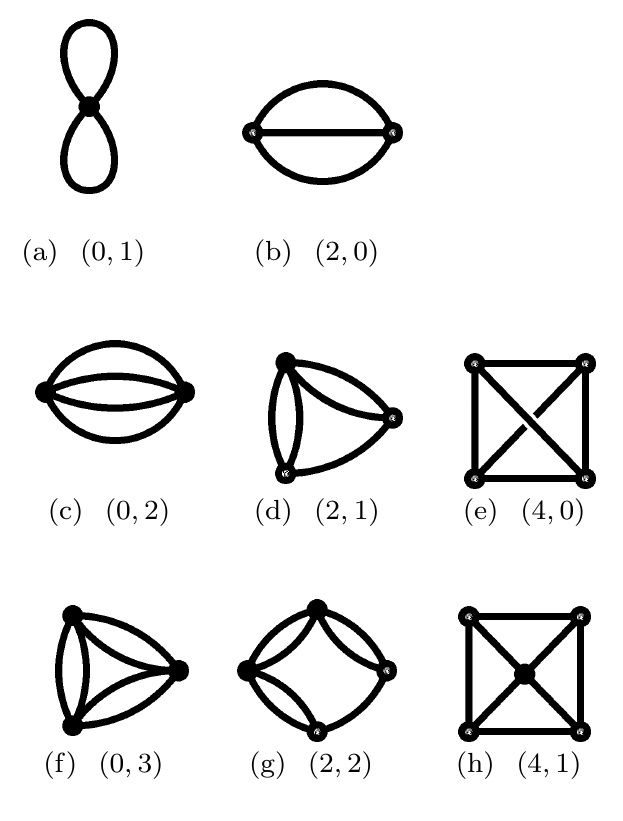}
\caption{Free energy diagrams, illustrating Table I. Each column has same number of triple vertices. Each row has same power of $T/J$.} \label{FDiagrams}
\end{figure}

We write the free energy density of the Garel-Doniach model as the Baym-Kadanoff functional~[\onlinecite{LuttingerWard,BaymKadanoff,Cornwall}], depending on the fully dressed Green's function. This functional is expanded perturbatively in powers of the quartic and the triple spin-wave interactions. In units of $T$, the Baym-Kadanoff functional can be conveniently arranged in terms with a growing number of triple vertices:
\begin{eqnarray}\label{OmegaQ}
2 \Omega[\mathbf{G}]/T&=&\int\left(\ln\frac{{\cal G}_0(\mathbf{p})}{\mathbf{G}(\mathbf{p})}
+\frac{\mathbf{G}(\mathbf{p})}{{\cal G}_0(\mathbf{p})} -1\right)
\frac{d^2\mathbf{p}}{(2\pi)^2}\nonumber \\
&&+\frac{1}{4t} (Q^2-\theta^2)^2 +\frac{3}{2} (Q^2-\theta^2) A_0[\mathbf{G}] \nonumber \\
&&+\sum_{k=0}^\infty Q^{2k} \Omega_{k}[\mathbf{G}],
\end{eqnarray}
where $\mathbf{G}(\mathbf{p})$ is the variational fully dressed Green's function. The functionals $\Omega_0[\mathbf{G}]$, $\Omega_1[\mathbf{G}]$ and $\Omega_2[\mathbf{G}]$ represent a sum of all the diagrams in the first, second and third column of Table~I correspondingly. The few first diagrams in each column are also shown in columns in Fig.~\ref{FDiagrams}. A diagram with a number $m$ of triple and a number $n$ of quartic vertices carries a coefficient:
\begin{equation}\label{Fmn}
(-1)^{n+1}t^{m/2+n}Q^{m},
\end{equation}
in front. The number of the triple vertices is always even. 
In those thermodynamic phases where the pitch field is absent, 
$\langle Q\rangle=0$, only quartic vertices do contribute to the free energy. In each box of Table~I, the combinatorial coefficient, a rational number, of the diagram is also shown, while inside the parentheses, the two numbers separated by comma are the numbers of the triple and the quartic vertices in the diagram, respectively. In higher orders of $Q$ and $T$, many diagrams do enter one box of the extended Table~I.

There is a remarkable relationship between the combinatorial coefficients in the first two columns of this table. Each diagram with just two triple vertices can be drawn from the diagram that includes only quartic vertices by cutting out one Green's function. This can be done by a number of ways that is equal to the number of the Green's functions. Therefore the combinatorial coefficient in a given box of the second column of Table~I equals to the combinatorial coefficient in the first column, one step down, by multiplying it by the number of the Green's functions.

We will need the few first terms of the expansion of the Baym-Kadanoff functional in powers of the normalized temperature $t$ explicitly:
\begin{equation}\label{Omega0}
\Omega_0[\mathbf{G}]=\frac{3t}{4}A_0[\mathbf{G}]^2 -\frac{3t^2}{8}B_2[\mathbf{G}] +\frac{9t^3}{8}B_3[\mathbf{G}] - \cdots
\end{equation}
and
\begin{equation}\label{Omega1}
\Omega_1[\mathbf{G}]=-\frac{3t}{2}A_1[\mathbf{G}] +\frac{27t^2}{4}A_2[\mathbf{G}] - \cdots .
\end{equation}
Each functional denoted by the capital letters $A$ and $B$ represents a contribution
from just one diagram, shown in Fig.~\ref{FDiagrams}. They are conveniently defined using the polarization operator functional:
\begin{equation}\label{POG}
\Pi[\mathbf{p},\mathbf{G}]= \int k_x^2 (p_x+k_x)^2 \mathbf{G}(\mathbf{k}) \mathbf{G}(\mathbf{p}+\mathbf{k}) \frac{d^2\mathbf{k}}{(2\pi)^2},
\end{equation}
in terms of the two sequences. The first sequence $n=0,\dots,\infty$ is
\begin{equation}\label{An}
A_n[\mathbf{G}]=\int p_x^2 \mathbf{G}(\mathbf{p}) 
\Pi^n[\mathbf{p},\mathbf{G}] \frac{d^2\mathbf{p}}{(2\pi)^2},
\end{equation}
and the second sequence $n=2,\dots,\infty$ is
\begin{equation}\label{Bn}
B_n[\mathbf{G}]=\int \Pi^n[\mathbf{p},\mathbf{G}] \frac{d^2\mathbf{p}}{(2\pi)^2}.
\end{equation}
There are, of course, many other diagrams, some of them being shown in the third column 
in Fig.~\ref{FDiagrams}, with the temperature order larger than four, that also contribute to the Baym-Kadanoff functional, Eq.~(\ref{OmegaQ}), but who are not members of these two sequences. We will account for them later in an approximate way.

The condition that the variation of the Baym-Kadanoff functional with respect to the fully dressed Green's function is stationary:
\begin{equation}\label{DySon}
\frac{\delta \Omega[\mathbf{G}]}{\delta\mathbf{G}}=0,
\end{equation}
gives the Dyson equation:
\begin{equation}\label{Dyson}
\mathbf{G}^{-1}(\mathbf{p})=G_0^{-1}(\mathbf{p})-\Sigma(\mathbf{p}).
\end{equation}
The variation of the first three terms in the functional~(\ref{OmegaQ}) gives
$-\mathbf{G}^{-1}(\mathbf{p})+{\cal G}_0^{-1}(\mathbf{p})+\frac{3}{2} (Q^2-\theta^2) p_x^2
\equiv -\mathbf{G}^{-1}(\mathbf{p})+G_0^{-1}(\mathbf{p})$, according to the definition of ${\cal G}_0(\mathbf{p})$ and $G_0(\mathbf{p})$ in Eqs.~(\ref{G0pQT0}) and (\ref{G0pQ}). Now, by varying the Baym-Kadanoff functional~(\ref{OmegaQ}), we find the expansion of the self-energy:
\begin{equation}\label{SelfEnergy}
\Sigma(\mathbf{p})=p_x^2\left(-\frac{3t}{2} A_0[\mathbf{G}] +\frac{9t Q^2}{2} \Pi[\mathbf{p},\mathbf{G}] + \cdots \right).
\end{equation}
The self-energy $\Sigma({\mathbf p})$ is proportional to $p_x^2$ in all orders. For the functional $\Omega_0[\mathbf{G}]$ this dependence on the momentum is exact, whereas the triple-vertex diagrams generate an additional dependence on the momentum $\mathbf{p}$. The self-energy contains many more diagrams, coming from combining the two vertices: the quartic $(\partial_x\psi)^4$ and the triple $\left(\partial_x\psi\right)^3$, in the Hamiltonian~(\ref{HpsiQ}), corresponding to the quartic and the triple spin-wave interactions. By construction the Baym-Kadanoff functional provides the same fully dressed Green's function as in the usual perturbation expansion in powers of the interaction Hamiltonian ${\cal H}_{int}$:
\begin{eqnarray}\label{HintT}
\frac{T}{J}&&{\bf G}({\bf r},{\bf r}')=\left\langle\psi({\bf r})\psi({\bf r}')\right\rangle
\nonumber \\
&&=\frac{\int {\cal D}[\psi]
\psi({\bf r})\psi({\bf r}') \, e^{-{\cal H}_0/T} 
\, e^{-{\cal H}_{int}/T}}{\int {\cal D}[\psi] \, e^{-{\cal H}/T}}\nonumber \\
&&=\frac{\int {\cal D}[\psi]\,
\psi({\bf r})\psi({\bf r}') \, e^{-{\cal H}_0/T} 
\,\left(1-\frac{{\cal H}_{int}}{T}+\frac{1}{2!}
\frac{{\cal H}_{int}^2}{T^2}-\cdots\right) }{\int {\cal D}[\psi] \, e^{-{\cal H}/T}}.\nonumber\\
\end{eqnarray}
If the Baym-Kadanoff functional included all the diagrams in all orders, then it would take on the same value for many different trial Green's functions $\mathbf{G}(\mathbf{p})$ and therefore the free energy could be evaluated by using a quite arbitrary Green's function. However, if the perturbation expansion is only partial or approximate, then the minimum of the Baym-Kadanoff functional is attained on the Green's function that is close to the true Green's function of a given system.

Since the self-energy is proportional to $p_x^2$ for the Garel-Doniach model~[\onlinecite{Doniach}], we make the following ansatz for the trial fully dressed Green's function:
\begin{equation}\label{GpQ}
\mathbf{G}^{-1}(\mathbf{p})=r p_x^2 +p_y^2 +\frac{1}{4}
p_x^4,
\end{equation}
with an adjustable temperature-dependent variational parameter $r$, which reflects the strong anisotropy in $x$-direction. However, one has to remember, that the variational approach with a constant parameter $r$ 
does not work properly in the range of the highly developed fluctuations, whereas the standard renormalization group, developed in Sec.~\ref{RG}, works well. In particular, near the chiral phase transition, this parameter will acquire dependence on momenta: 
$$r=r(p_x,p_y),$$ and the short series Baym-Kadanoff mean-field functional formalism will effectively transform into renormalization group parquet diagrams. The real Green's function in the critical region has a quite different scaling dimensionality.

Using the ansatz~(\ref{GpQ}) for the fully dressed Green's function, we reduce the Baym-Kadanoff functional (\ref{OmegaQ}) to a simple function:
\begin{eqnarray}\label{OmegaQlambda}
\frac{2}{T} \Omega(r,Q)&\!=\!&\frac{1}{\pi}\left(\frac{2}{3}r^{3/2}-2 \theta^2 \sqrt{r}\right)
\nonumber \\
&&+\,\frac{1}{4t} (Q^2-\theta^2)^2 +\frac{3}{2} (Q^2-\theta^2) A_0(r) 
\nonumber \\
&&+\left(\frac{3t}{4} A_0(r)^2 -\frac{3t^2}{8} B_2(r)+\frac{9t^3}{8} B_3(r) - \cdots \!\right) \nonumber \\
&&+\,Q^2 \!\left( -\frac{3t}{2} A_1(r)+\frac{27t^2}{4} A_2(r)+ \cdots \right),\nonumber \\
\end{eqnarray}
of the two parameters $r$ and $Q$, the first representing the spin stiffness and the second representing the effective helical structure pitch. In the minimum of the Baym-Kadanoff functional, they both become temperature-dependent. Here the free energy satisfies the two stationary conditions:
\begin{equation}\label{stationary}
\frac{\partial \Omega}{\partial Q}=0, \quad \frac{\partial \Omega}{\partial r}=0.
\end{equation}
At zero temperature, the solution of the stationary conditions can be read off directly from the Hamiltonian~(\ref{HpsiQ}) of the Garel-Doniach model:
\begin{equation}
r=\theta^2, \quad Q=\theta,
\end{equation}
where $\theta$ is the helicity of the magnet.

Let us now prove, that when the temperature is increased, the stationary conditions~(\ref{stationary}) allow for a special solution:
\begin{equation}\label{SolH}
r=0, \quad Q=0.
\end{equation}
From outright, we neglect the diagrams in the third and further columns on the right side of Table~I. Using Eq.~(\ref{OmegaQlambda}), we find
\begin{equation}\label{SE1}
0=\frac{\partial \Omega}{\partial Q^2}=-\theta^2 + 3t A_0 -3t^2 A_1 +\frac{27 t^3}{2} A_2 - \cdots,
\end{equation}
at $Q=0$, with only the second column of diagrams present. The second stationary condition in 
Eq.~(\ref{stationary}) at $Q=0$ involves only diagrams from the first column of Table~I. Differentiating with respect to $r$ means selecting one Green's function and differentiating it. 
Using the definition Eqs.~(\ref{POG})-(\ref{Bn}) and using the Green's function Eq.~(\ref{GpQ}), we derive the following relationships:
\begin{equation}
\frac{\partial A_0(r)}{\partial r}= -\frac{1}{\pi\sqrt{r}}, \quad \frac{\partial B_k(r)}{\partial r}= -2 k\frac{1}{\pi\sqrt{r}} A_{k-1}(r),
\end{equation}
for $k\geq 2$ and small $r$, that can be checked by direct inspection. The coefficient $2k$ equals to the number of the Green's functions and we find that under differentiation both the content of the diagram and the combinatorial coefficient are transferred from the first column to the second one. Thus we find the second stationary equation for the free energy, Eq.~(\ref{OmegaQlambda}):
\begin{equation}\label{SE2}
0=-2r-\theta^2 + 3t A_0 -3t^2 A_1 +\frac{27 t^3}{2} A_2 - \cdots,
\end{equation}
at $Q=0$. We observe that the sequence of diagrams here is exactly the same as in the first stationary equation~(\ref{SE1}). Therefore the two stationary equations~(\ref{SE1}) and (\ref{SE2}) have a solution~(\ref{SolH}). This solution, $r=0$ and $Q=0$, represents a chiral phase transition from a paramagnetic state at high temperatures to a chiral spin liquid state with $Q\neq 0$ at low temperatures. 

The one-loop polarization operator is a complicated function of the transferred momentum $\mathbf{p}$. In the low-temperature limit, though, it becomes scale invariant. This means that under the scale transformation:
\begin{equation}
p_x\to \lambda p_x, \quad p_y\to \lambda^2 p_y, \quad r \to \lambda^2 r,
\end{equation}
the main part of the polarization operator transforms like
\begin{equation}
\Pi_0(\mathbf{p}) \to \frac{1}{\lambda} \Pi_0(\mathbf{p}).
\end{equation}
In addition to this leading term there are next-order terms, controlled by the power of the normalized temperature $t$, coming from the effects of the finite Brillouin zone. In the following, we will limit our analysis to the scale invariant part of the polarization operator and omit the index zero. At vanishing momentum, we find
\begin{equation}
\Pi(\mathbf{p}=0)=\frac{1}{\pi \sqrt{r}}.
\end{equation}
However, we will need the polarization operator more in the limit $|\mathbf{p}|\gg \sqrt{r}$. In this case, using the definition Eq.~(\ref{POG}), the Green's function ansatz~(\ref{GpQ}) and taking the integral with respect to the internal momentum, we find
\begin{eqnarray}
\Pi(\mathbf{p})=\frac{\sqrt{2\sqrt{p_x^4 + 16 p_y^2}+2p_x^2}}{\sqrt{p_x^4+16 p_y^2 }} -\frac{4}{\pi}p^2_x\sqrt{r} \mathbf{G}(\mathbf{p}) + O(r),\nonumber \\
\end{eqnarray}
where the first term is the exact value of the polarization operator at $r=0$, whereas the next terms are series expansion in small $r$.

To proceed further, we need an approximation to the infinite series of diagrams in the first and second columns of Table~I. Since we are developing a method suited for the low temperatures, we will try to keep the three lowest order diagrams in the sequence exactly, whereas for the sum of all higher order diagrams, we impose only a physical bound that they do not exceed by far the value of the low order diagrams as the temperature rises, $T\to\infty$. These conditions can be met by using the so-called Pad\'{e} approximation. For instance, the sum of all the diagrams in the second column of Table~I is given by the following Pad\'{e} approximation:
\begin{equation}
\Omega_1(r)=-\frac{3}{2}\int \frac{t\Pi(\mathbf{p})}{1+\frac{9}{2}t\Pi(\mathbf{p})} p_x^2\mathbf{G}(\mathbf{p}) \frac{d^2\mathbf{p}}{(2\pi)^2}.
\end{equation}
Thus, using the Pad\'{e} approximation, we can write the stationary equations in a closed form. The first stationary equation (\ref{stationary}) represents the condition of the stability of the helical state:
\begin{eqnarray}\label{forQ}
Q\left( Q^2-\theta^2+3t \int \frac{1+\frac{7}{2}t\Pi(\mathbf{p})}
{1+\frac{9}{2}t\Pi(\mathbf{p})} p_x^2\mathbf{G}(\mathbf{p}) \frac{d^2\mathbf{p}}{(2\pi)^2} \right) = 0,\quad\quad
\end{eqnarray}
neglecting the term of the order $t^3Q^2$, coming from the higher order diagrams in the third and higher columns. It always has one solution $Q=0$. This stability condition reflects that at equilibrium the system chooses between the two solutions: $\pm |Q|\neq 0$ in the low-temperature region and $Q=0$ at and above the chiral phase transition.

Next, we have to minimize the free energy with respect to the spin stiffness $r$: $\partial\Omega/\partial r=0$, which determines $r$ as a function of the order parameter $Q$:
\begin{eqnarray}\label{forlambda}
0\!&=\!&\frac{1}{\pi}\frac{r-\theta^2}{\sqrt{r}}-\frac{3}{2}Q^2\frac{\partial}{\partial r} \int \frac{t\Pi(\mathbf{p})}{1+\frac{9}{2}t\Pi(\mathbf{p})} p_x^2 \mathbf{G}(\mathbf{p}) \frac{d^2\mathbf{p}}{(2\pi)^2}\nonumber\\
&&+\frac{3}{2}\left(Q^2-\theta^2 + t\int \frac{1+\frac{7}{2}t\Pi(\mathbf{p})}
{1+\frac{9}{2}t\Pi(\mathbf{p})} p_x^2\mathbf{G}(\mathbf{p}) \frac{d^2\mathbf{p}}{(2\pi)^2}  \! \right)\frac{\partial A_0}{\partial r}.\nonumber\\
\end{eqnarray}
We find the low-temperature asymptotics of the $\Omega_1(r)$ functional of the free energy:
\begin{eqnarray}
\int \frac{t\Pi(\mathbf{p})}{1+\frac{9}{2}t\Pi(\mathbf{p})} p_x^2\mathbf{G}(\mathbf{p}) \frac{d^2\mathbf{p}}{(2\pi)^2}\approx  \frac{2}{\sqrt{3}\pi}\ t \ln\frac{1}{t} -\frac{1}{\pi} \sqrt{r}\qquad\quad
\end{eqnarray}
and also 
\begin{equation}
A_0(r)\approx \frac{1}{\pi} \left(2.30218 - 2\sqrt{r}\right),
\end{equation}
where the upper momentum cut-off constant comes from limiting the integration to be within the Brillouin zone: $p_{x,y}\in(-\pi,\pi)$. The system of equations~(\ref{forQ}) and (\ref{forlambda}) provides the mean-field solutions for the pitch-field order parameter $Q$ and the spin stiffness $r$ in a wide region of temperatures, which includes both regions below and above the chiral phase transition. We find from Eq.~(\ref{forQ}) the position of the chiral phase transition line in the plane $(t,\theta)$:
\begin{equation}\label{asymptoticH}
\theta_{c}^2(t)=\frac{3t}{\pi}\left(2.30218- \frac{2}{\sqrt{3}} \ t\ln\frac{1}{t} \right),
\end{equation}
asymptotically for the small normalized temperature $t=T/J$.

In the end of this section, we note that the small deviations of the pitch field $Q$ from the equilibrium value can be described by the Ginsburg-Landau functional. Indeed, eliminating the variational spin stiffness $r$, we obtain the free energy functional expanded in powers of the order parameter, the pitch-field $Q$, for any given thermodynamic pair $(T,\theta)$:
\begin{equation}
\Omega[Q]=\Omega_0+\int \left( \frac{1}{2} m^2 Q^2 +\frac{1}{4!}  g Q^4 + \cdots\right)\ d^2\mathbf{r},
\end{equation}
where in the first term $\Omega_0$ all the short-wavelength fluctuations (of the field $\psi$) are summed up. The condition $m^2(T_c)=0$, $g>0$ determines the second-order chiral phase transition line. We will pursue the theory in the critical region along these lines in Sec.~\ref{RG}.

\section{Phase diagram}
\label{SectionPhaseDiagram}

In the long-range limit the correlation functions of the magnetization are determined by the long-range asymptotics of the Green's function:
\begin{equation}
\mathbf{G}(\mathbf{p})=\frac{1}{r p_x^2 +p_y^2}.
\end{equation}
The nonlinear terms from the higher order diagrams in the expansion in quartic and triple vertices contribute to the local core energy of the vortex. In the vicinity of the chiral phase transition line, where $r\approx 0$, the interaction between the vortices is short range~[\onlinecite{LNP}]. The Berezinskii-Kosterlitz-Thouless phase transition~[\onlinecite{B,KT,K}] is due to the dissociation of the vortex-antivortex pairs. The condition for the dissociation is the condition of the nulling of the vortex free energy:
\begin{eqnarray}\label{KT}
F_v=\left(\pi J \sqrt{r(T_{BKT})}-2 T_{BKT}\right) \ln \frac{L}{a}=0,
\end{eqnarray}
which gives for the Berezinskii-Kosterlitz-Thouless critical temperature:
\begin{eqnarray}\label{defTBKT}
T_{BKT}=\frac{\pi J}{2}\sqrt{r(T_{BKT})},
\end{eqnarray}
or, equivalently, the critical condition: $\sqrt{r}=2t/\pi$.

\begin{figure} 
\includegraphics[width=0.478\textwidth]{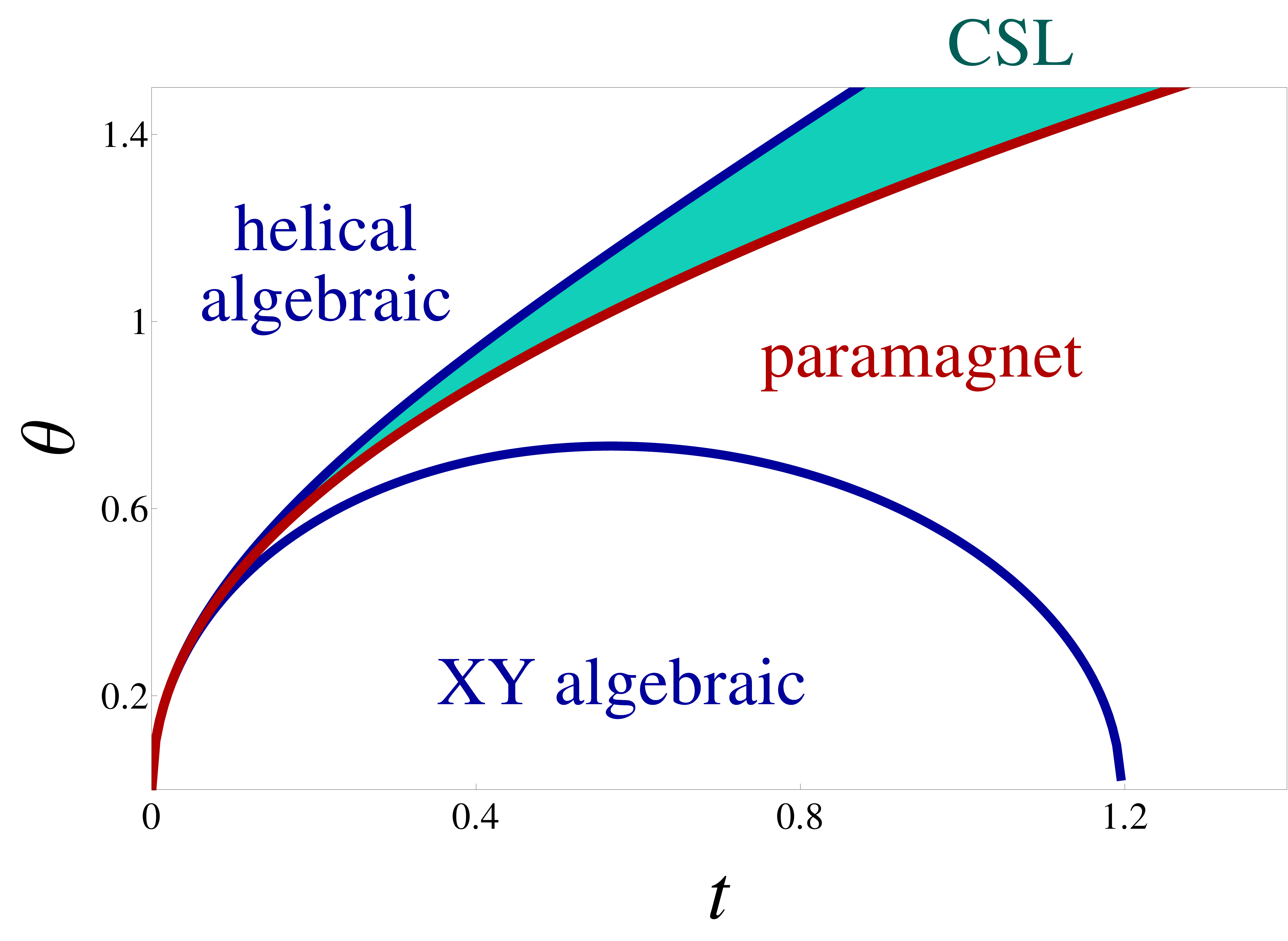}
\caption{(Color online) Red, in the center, is the line of the second-order chiral phase transition. The two blue lines, surrounding it, are the Berezinskii-Kosterlitz-Thouless phase transitions. CSL stands for the chiral spin liquid phase. } \label{phasediagram}
\end{figure}

Now, we obtain the three critical lines of the three continuous phase transitions on the phase diagram $(t,\theta)$, selecting three of the following equations:
\begin{eqnarray}
&&Q=0, \quad r=0, \quad r=\frac{4}{\pi^2} t^2, \quad 2r= \frac{1}{2} Q^2, \nonumber\\
&&\theta^2=\frac{3}{2}Q^2-2r+3t \int \frac{1+\frac{7}{2}t\Pi(\mathbf{p})}
{1+\frac{9}{2}t\Pi(\mathbf{p})} 
p_x^2\mathbf{G}(\mathbf{p}) \frac{d^2\mathbf{p}}{(2\pi)^2}.\quad\quad
\end{eqnarray}
The first, second, and the last one determine the chiral phase transition line. The first, third, and the last one determine the Berezinskii-Kosterlitz-Thouless phase transition line above the chiral phase transition line. The three last equations together determine the Berezinskii-Kosterlitz-Thouless phase transition line below the chiral phase transition line. We evaluate the integral numerically at all temperatures and find the phase diagram, shown in Fig.~\ref{phasediagram}. In the same way as the asymptotic chiral phase transition line was found in Eq.~(\ref{asymptoticH}), we find for the two Berezinskii-Kosterlitz-Thouless phase transition lines:
\begin{equation}\label{asymptoticBKTg}
\theta_+^2(t)=\theta_{c}^2(t)-\frac{10}{\pi^2}t^2
\end{equation}
and
\begin{equation}\label{asymptoticBKT}
\theta_-^2(t)=\theta_{c}^2(t)+\frac{2}{\pi^2}t^2,
\end{equation}
asymptotically for the small $t=T/J$.

We plot the phase diagram in the plane of the temperature and the helicity parameter: $(t,\theta)$, shown in Fig.~\ref{phasediagram}. At the origin $(0,0)$, we find the Lifshitz point, where all the three phase transition lines start, and in particular the second order chiral phase transition line, Eq.~(\ref{asymptoticH}), starts. The average pitch-field order parameter vanishes on this line: $\langle Q\rangle =0$. The two lines of the Berezinskii-Kosterlitz-Thouless phase transitions 
Eqs.~(\ref{asymptoticBKTg}) and (\ref{asymptoticBKT}) surround the chiral phase transition. One of the two magnetically ordered phases extends into the high temperatures, where our method provides results only qualitatively.  

In two dimensions, there is no long-range order in systems possessing a continuous symmetry. Yet, for the two-component magnet with a continuous $O(2)$ symmetry, the so-called XY magnet, there exists a nonlocal order parameter, defined as a correlation function of the magnetization at two distant points. At a low temperature, where vortices are bound in pairs, it decays weakly as a power-law function:
\begin{equation}
\left\langle \vec{m}(\mathbf{x})\cdot \vec{m}(\mathbf{y})  \right\rangle= 
\left\langle e^{i \left( \phi(\mathbf{x})- \phi(\mathbf{y}) \right)} \right\rangle \sim \frac{1}{|\mathbf{x}-\mathbf{y}|^{2\Delta_\phi(T)}},
\end{equation}
with an infinite correlation length, whereas at high temperatures, when free vortices proliferate, it decays fast as an exponential function with a finite correlation length. For systems with such a nonlocal order parameter, a natural question arises. Imagine that a second order parameter emerges in the system, serving as a gauge connection to the fields in the first, nonlocal order parameter. For example, in an XY magnet, a local pitch-field order $Q(\mathbf{r})$ can develop, when parity is broken in the low-temperature state of this helical magnet. The idea to consider the spin chirality, connected to the pitch of the helical ground state, as the order parameter rather than the spin variable belongs to Villain~[\onlinecite{Villain}]. In a parametrization of the XY magnetization via the phase $\phi(\mathbf{r})$, the nonlocal order parameter and the gauge connection due to the pitch field are exactly equal:
\begin{equation}
\phi(\mathbf{r})-\phi(\mathbf{r'})=\int_\mathbf{r'}^\mathbf{r} \vec{\partial}\phi \cdot d\vec{l}.
\end{equation}
However, for systems in the vicinity of the phase transitions of the second order, the order parameter, like the pitch-field, strongly fluctuates and is in fact an average over a large spin block. This renders the above equality invalid. A new gauge transformed, in a rotating frame, nonlocal order parameter
\begin{eqnarray}\label{guageOrder}
&&\left\langle R^{\alpha\beta}_{\langle Q\rangle} m^\beta(\mathbf{x}) \cdot R^{\alpha\gamma}_{\langle Q\rangle} m^\gamma(\mathbf{x'})  \right\rangle \\
&&\qquad=\left\langle \cos\left( \phi(\mathbf{r})-\phi(\mathbf{r'}) 
- \int_\mathbf{r'}^\mathbf{r} \mathbf{Q}(\mathbf{l})\cdot d\mathbf{l} \right) \right\rangle
\nonumber 
\end{eqnarray}
may define a new thermodynamic phase. The physical meaning of this phase can be understood if the local order parameter $Q$ is one-component, representing a discrete $Z_2$ Ising order. In the ordered phase, the local pitch-field order parameter takes on two opposite values: the first inside a percolating majority domain, $Q(\mathbf{r})=+Q_0$, and the second inside a few minority domain inclusions, $Q(\mathbf{r})=-Q_0$. For the purpose of a gauge connection, in this case, there exist two gauge fields: the average $\langle Q\rangle$ and $Q_0$. Let us also augment the definition of the nonlocal 
order parameter, Eq.~(\ref{guageOrder}), by averaging it around points $\mathbf{r}$ and $\mathbf{r'}$ by an area larger than the pitch step $2\pi/Q$. Then, if the gauge $Q_0$ transformed nonlocal order parameter is non-zero and the gauge $\langle Q\rangle$ transformed nonlocal order parameter is zero, then domain walls in the $Z_2$ pitch-field order are present in the system.

The results, that have been obtained until now for the
ferromagnetic (FM) XY frustrated classical model, can be
readily extended to the antiferromagnetic (AF) XY model,
with a reversed sign of the nearest-neighbor interactions. For
this purpose, we superimpose a sublattice with a black and
white chequered board order on the square lattice, placing
each next-nearest neighbor spin on a same color square of
the chessboard. Further, we flip all the spins on the black
chessboard squares, and leave the same all the spins on the
white chessboard squares. Such a modification is equivalent to
a sign reversal of all the nearest-neighbor interactions $J_1$ in the
Hamiltonian~(\ref{Hamiltonian}), while leaving the sign of the next-nearest
antiferromagnetic interactions $J_2$ the same. The spins are classical, thus no
insertion of spin commutators in the Hamiltonian accompanies
the spin flips. This means, that there is a one-to-one mapping
of the classical FM onto the classical AF model, which turn out
to be identical. A corollary on the phase diagram of the helical
magnet is that all the three phase transition lines, obtained for
the FM model at values of the helicity parameter close to zero,
$\theta\to 0$, shown in Fig.~\ref{phasediagram}, will be repeated 
for the AF model, at $\theta\to \pi$, via a mirror transformation 
around the axis $\theta=\pi/2$ in the plane $(t,\theta)$.
As a consequence, for instance, the two critical points above which 
the FM and the AF states, existing in the regions of $\theta\to 0$ and 
$\theta\to\pi$ respectively, become paramagnetic, 
are equal: $T_{Curie}=T_{Neel}$. At $J_1=0$, or $\theta=\pi/2$,
the Hamiltonian~(\ref{Hamiltonian}) reduces to that of a one-dimensional
antiferromagnet, which has no helical state, but undergoes
a Kosterlitz-Thouless transition at a nonzero temperature.
This implies that at some finite large value of the helicity
parameter $\theta\in (0;\pi/2)$, beyond the scope of Fig.~\ref{phasediagram}, 
the chiral phase transition line and the inner Kosterlitz-Thouless phase
transition line should intersect. However, studying a magnet
with such a large helicity, when the continuum approach is
no longer valid, is beyond the scope of the present paper
and demands developing other methods. The classical model,
considered in this paper, can describe real physical systems
with large spins. In analogous quantum AF and FM models,
the dispersion laws differ: in the AF problem zero quantum
fluctuations appear as a result of the frustration, while the FM
problem lacks them. Thus contrary to the classical case, there
is no one-to-one mapping between the quantum AF and the
quantum FM models of the helical magnet.

\section{Critical behavior at the chiral phase transition}
\label{RG}

In Sec.~\ref{Mean-field}, we have found the Landau mean-field free energy in terms of the uniform pitch-field order parameter $Q$. Since we will rely heavily on the universality, the gradient terms can be taken from the Hamiltonian, neglecting the entropy corrections. We introduce an auxiliary field $\psi$, not to be confused with the field in the rotating frame in Sec.~\ref{Model}, to deal with the apparent nonlocality of the energy in terms of the pitch-field $Q(\mathbf{r})$. Using the identity
\begin{equation}
e^{
-\frac{1}{2} \int (\partial_y\phi)^2 d\mathbf{r}} 
\! = \!\int \mathcal{D}\psi e^{
-\int \left[\frac{1}{2}(\partial_x\psi)^2 +i \partial_x\phi \partial_y\psi \right] d\mathbf{r}},
\end{equation}
we find the local free energy functional for the spatial variations of the pitch-field $Q(\mathbf{r})=\partial_x\phi(\mathbf{r})$:
\begin{eqnarray}\label{action}
\Omega[Q]&=&\int \left[\frac{1}{2}(\partial_x Q)^2+ \frac{1}{2} m^2 Q^2 +\frac{1}{4!}  g Q^4\right. \nonumber \\
&&\left.+\frac{1}{2}(\partial_x \psi)^2 +iQ \partial_y \psi -h Q \right]\ d^2\mathbf{r}.
\end{eqnarray}
There is no trace of the magnetization vector $\mathbf{m}(\mathbf{r})$ in this description. Vortices~[\onlinecite{LNP}], abundant in the vicinity of the chiral phase transition line, destroy the magnetization on the short scale. The description of the critical model in terms of the local pitch field $Q(\mathbf{r})$, which is decoupled from its low-temperature definition:
\begin{equation}
Q(\mathbf{r})=\langle [\mathbf{m}\times \partial_x\mathbf{m}] \rangle\cdot \mathbf{n},
\end{equation}
where $\mathbf{n}$ is a normal to the plane, 
due to abundant vortices, is therefore appropriate from the physical point of view. The field  $h$ in 
Eq.~(\ref{action}) conjugated to the order parameter $Q$ is a twist field, associated with the change of the boundary condition in such a way as to induce a pitch in the bulk. A true magnetic field couples to the local magnetization which is a nonlocal operator in the representation Eq.~(\ref{action}).

We will study the critical phenomena for the above model in a higher dimension $d=2-\epsilon$, 
replicating the $x$ coordinate up to two ones: $dyd^d\mathbf{x}$. In the same time, we will keep 
$Q(y,\mathbf{x})$ as a scalar one-component fluctuating field. 
In order to make a dimensional regularization and in order to develop an $\epsilon$-expansion renormalization, we need to symmetrize the mass-term:
\begin{eqnarray}
\Omega[Q]&=&\int \left[ \frac{1}{2}(\nabla Q)^2+ \frac{1}{4} m^2 Q^2 +\frac{1}{4!}  g Q^4
\right.\nonumber \\
&&+\left.\frac{1}{2}(\nabla \psi)^2 +iQ \partial_y \psi +\frac{1}{4} m^2 \psi^2 -h Q\right]\ dyd^d\mathbf{x},\nonumber \\
\end{eqnarray}
where $\nabla=\partial_{\mathbf{x}}$. Anyway, the mass term will be forcibly kept at zero exactly at the critical point. In two dimensions, an important question is whether vortices have an effect on the helical phase transition. A vortex is generated by the duality transformed field operator:
\begin{equation}
\cos\left( m\int \partial_x \phi \ dy\right)= \cos\left( m\int Q \ dy\right).
\end{equation}
The scaling dimension of the argument of the $\cos$ function is zero, while $m\to 0$ at the critical point. Therefore we conclude that the vortices are unbound, free at the chiral phase transition.

We proceed using the conventional renormalization group method, that is explained in details in Appendix. At the chiral phase transition, we find the following scaling relationships:
\begin{eqnarray}\label{scaling}
\beta\delta &=&\beta+\gamma, \quad
\alpha+2\beta+\gamma = 2, \nonumber  \\ 
\\(2-\eta)\nu &=&\gamma,  \quad
\left(d+2-{\textstyle \frac{1}{2}}\eta\right)\nu = 2-\alpha,  \nonumber
\end{eqnarray}
connecting the chiral critical exponents. $\gamma$ is the chiral critical exponent for the twist-susceptibility. The last relationship differs from the usual Josephson scaling relationship: $d\nu=2-\alpha$. This deviation is due to the fact that in the chiral phase transition critical point there is no full conformal symmetry, since the spatial coordinates are nonequivalent. A reduced conformal symmetry in the higher dimensions space $\mathbf{x}$ probably holds. In our case of the two-dimensional helical magnet, when there is only one $x$ coordinate, this remnant conformal symmetry will degenerate.

The chiral critical exponents are found in Appendix in terms of the $\epsilon$ expansion as follows:
\begin{eqnarray}\label{critexp}
&&\nu=\frac{1}{2}+\frac{1}{12}\epsilon+\frac{1}{36}
\left(\ln{\frac{4}{3}}+\frac{67}{54}\right)\epsilon^2,\nonumber \\
&&\gamma=1+\frac{1}{6}\epsilon+\frac{1}{18}
\left(\ln{\frac{4}{3}}+\frac{59}{54}\right)\epsilon^2
,\nonumber \\
&&\beta=\frac{1}{2}-\frac{1}{6}\epsilon
+\frac{1}{36}
\left(\ln{\frac{4}{3}}-\frac{5}{27}\right)\epsilon^2,\nonumber \\
&&\delta=3+\epsilon+\frac{77}{162}\epsilon^2,
\nonumber \\
&&\alpha=\frac{1}{6}\epsilon-\frac{1}{9}
\left(\ln{\frac{4}{3}}+\frac{49}{108}\right)\epsilon^2,
\end{eqnarray}
and
\begin{equation}\label{eta}
\eta=\frac{4}{243} \epsilon^2 - \frac{4}{19683}\left(94+108\ln 2 -135\ln 3\right) \epsilon^3.
\end{equation}
They differ from the critical exponents of all known $O(N)$ phase transitions of the second order in magnets with local exchange interactions. The chiral critical exponents, Eq.~(\ref{critexp}), have been found~[\onlinecite{BrezinZinnJustin}] in the Ising model 
with strong long-range dipolar interactions~[\onlinecite{Larkin},\onlinecite{Aharony}], while Eq.~(\ref{eta}) extends the known results. Therefore we conclude this section having found the equivalence of the universality class of the chiral phase transition of the second order from the paramagnet into the chiral spin liquid and the universality class of the two-dimensional Ising model 
with strong long-range dipolar interactions~[\onlinecite{Larkin},\onlinecite{Aharony}].

\section{Discussion and Conclusions}
\label{Discussion}

The phase diagram of the Garel-Doniach model is found analytically in an asymptotically exact way at low temperatures. One line of the chiral phase transition is surrounded by the two lines of the Berezinskii-Kosterlitz-Thouless magnetic phase transitions. The universality class of the chiral phase transition is determined and turns out to be that of the two-dimensional Ising model 
with strong long-range dipolar interactions. The existence of the chiral spin liquid phase is proven analytically at low temperatures. We find that the two-dimensional Garel-Doniach model is a rare example of a simple exchange magnet on a lattice, where the spin liquid phase does develop. 
As it breaks the parity, this spin liquid is a chiral spin liquid.

Our method applies to a classical helical magnet only, i.e., the one in the limit of a large on-site spin $S$. The effects of the quantum fluctuations when the on-site spin $S$ is small remain to be investigated. But one feature in our phase diagram, namely, that the spin liquid phase extends all the way down to zero temperature, is very interesting. The Lifshitz point, where the three lines of the phase transitions originate, will be the quantum phase transition point, extending its quantum criticality influence on the chiral spin liquid.

\section*{Acknowledgments} 
We thank H. Schenck, V. L. Pokrovsky and T. Nattermann for sharing details of their ongoing investigation of the same problem. 

\appendix*
\section{Renormalization group analysis}
\label{AppRG}

In this Appendix, we calculate the chiral critical exponents using dimensional regularization and the analysis of renormalization of the Ginsburg-Landau free energy functional in dimension $d=2-\epsilon$. We start with writing the renormalized free energy functional:
\begin{eqnarray}\label{rOmega}
\Omega_r \!&=\!\!&\int \!\left[ \frac{1}{2} Z(g) (\nabla Q)^2\!+ \!
\frac{1}{4} Z_m(g) m^2 Q^2\!+
iQ \partial_y \psi +\!\frac{1}{2}(\nabla \psi)^2 \right.\nonumber \\
&&+\!\left.
\frac{Z_m(g)}{Z(g)} \frac{1}{4} m^2 \psi^2\!+\!\frac{1}{4!} m^{2-d} g Z_g(g) Q^4\right]\ \!\!dyd^d\mathbf{x},
\end{eqnarray}
where we omit the index $r$ from the fields $Q_r\to Q$. The Green's function for the pitch-field reads, assuming the factor $Z_m=Z(g)$:
\begin{equation}
G(\mathbf{p},p_y)=
\frac{\mathbf{p}^2+\frac{1}{2} m^2}{Z(g) \left(\mathbf{p}^2+ \frac{1}{2} m^2\right)^2 +p_y^2}.
\end{equation}
The correction to the auxiliary field $\psi$ is found finite and it does not renormalize. 
We observe that the vertex $g$ enters the perturbation series always as a term:
\begin{equation}
G(g)=\frac{Z_g(g)}{Z^{3/2}(g)} g.
\end{equation}
Therefore the $\beta$ function is defined as follows:
\begin{equation}\label{beta}
\beta(g)=-(2-d)\frac{dg}{d\ln G(g)}.
\end{equation}
The two other multiplicative renormalization constants in the renormalized free energy, Eq.~(\ref{rOmega}), stand for the pitch-field: $Z(g)$, and for the so-called $Q^2$ insertion: $Z_2(g)$. We use the definition of the corresponding two critical exponents:
\begin{equation}\label{etaG}
\eta(g)=\beta(g)\frac{d}{dg}\ln Z(g)
\end{equation}
and
\begin{equation}\label{eta2G}
\eta_2(g)=\beta(g)\frac{d}{dg}\ln \frac{Z_2(g)}{Z(g)}.
\end{equation}

\begin{figure} 
\includegraphics[width=0.35\textwidth]{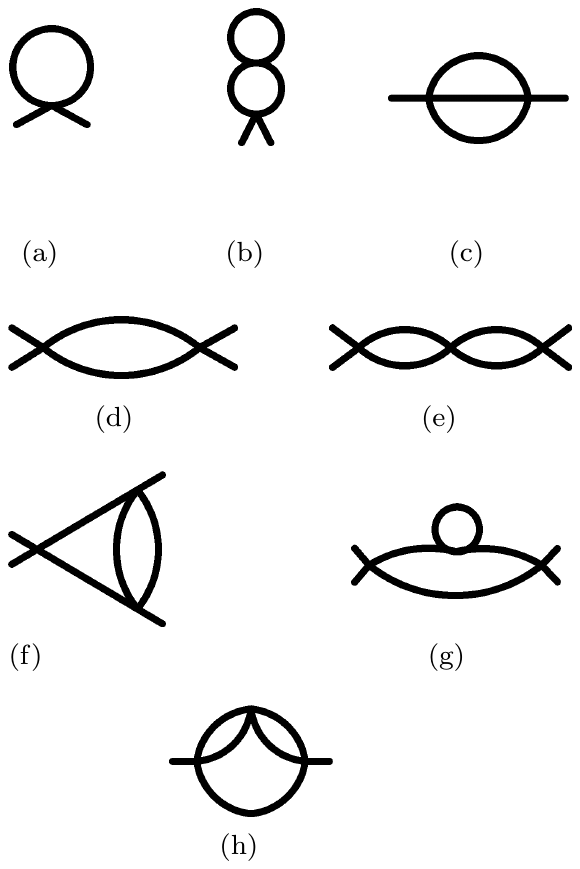}
\caption{Diagrams for the $\phi^4$-theory} \label{RGDiagrams}
\end{figure}

For the free energy, Eq.~(\ref{rOmega}), the renormalization proceeds similarly as in the usual 
$\phi^4$-theory and is given by the diagrams (a)-(h), shown in Fig.~\ref{RGDiagrams}. We find for instance the diagrams $(a)=0$, $(b)=0$. The diagram
\begin{eqnarray}
(d)&=&\int \frac{\left(\mathbf{p}^2+\frac{1}{2}m^2\right)^2}{\left(\left(\mathbf{p}^2+\frac{1}{2}m^2\right)^2+p_y^2\right)^2} \frac{dp_y}{2\pi} \frac{d^d\mathbf{p}}{(2\pi)^d} \nonumber \\
&=&\frac{1}{4} \int \frac{1}{\mathbf{p}^2+\frac{1}{2}m^2} \frac{d^d\mathbf{p}}{(2\pi)^d}
\\
&=&\frac{1}{8} \left(\frac{m^2}{2}\right)^{\frac{d}{2}-1} N_d \Gamma\left(\frac{d}{2}\right) \Gamma\left(1-\frac{d}{2}\right)\approx \frac{1}{4} N_d\frac{1}{\epsilon},
 \nonumber
\end{eqnarray}
where $N_d\approx 1/(2\pi)$. Next, the diagram $(e)=(d)^2$. The diagram (c) is
\begin{eqnarray}
(c)&=&\int \frac{dp_y}{2\pi} \frac{d^d\mathbf{p}}{(2\pi)^d} \frac{dq_y}{2\pi} \frac{d^d\mathbf{q}}{(2\pi)^d} G(\mathbf{p},p_y) G(\mathbf{q},q_y)
\nonumber\\
&&\times G(\mathbf{p}+\mathbf{q}+\mathbf{s},p_y+q_y+s_y),
\end{eqnarray}
where $\mathbf{s}$ is the transferred momentum. An expansion in $s_y$ gives a finite term, therefore, we set $s_y=0$. Integrating out $p_y$ and $q_y$:
\begin{eqnarray}
(c)&=&\frac{1}{4} \int \frac{d^d\mathbf{p}}{(2\pi)^d} \frac{d^d\mathbf{q}}{(2\pi)^d} 
\frac{1}{\mathbf{p}^2+\mathbf{q}^2+ (\mathbf{p}+\mathbf{q}+\mathbf{s})^2+\frac{3}{2}m^2} \nonumber\\
&=&\int_0^\infty \frac{dt}{4} \int \frac{d^d\mathbf{p}}{(2\pi)^d} \frac{d^d\mathbf{q}}{(2\pi)^d}e^{-t(\mathbf{p}^2+\mathbf{q}^2+ (\mathbf{p}+\mathbf{q}+\mathbf{s})^2+\frac{3}{2}m^2)}. \nonumber\\
\end{eqnarray}
We transform the term in the exponent as
\begin{equation}
(\mathbf{p},\mathbf{q})\left( \begin{array}{cc} 2t & t \\ t & 2t \end{array} \right) \left( \begin{array}{c} \mathbf{p} \\ \mathbf{q} \end{array} \right) + 2t\ (\mathbf{p},\mathbf{q})  \left( \begin{array}{c} \mathbf{s} \\ \mathbf{s} \end{array} \right) +t\mathbf{s}^2 + t \frac{3}{2}m^2.
\end{equation}
The determinant of this matrix is $3t^2$, inverting it and integrating the momenta, we find
\begin{eqnarray}\label{sunrise}
(c)=\frac{1}{16} N_d^2 \Gamma\left(\frac{d}{2}\right)^2 \! \int_0^\infty \! dt \frac{1}{(3t^2)^{d/2}} e^{
-t \frac{1}{3} \mathbf{s}^2-t\frac{3}{2}m^2}.\qquad\qquad
\end{eqnarray}
We expand the diagram (c) in the second power of the transferred momentum $\mathbf{s}$ and obtain the coefficient
\begin{eqnarray}
-\frac{1}{16} \frac{N_d^2}{3^{d/2+1}} \Gamma\left(\frac{d}{2}\right)^2 \Gamma(2-d) \left( \frac{3}{2}m^2\right)^{2-d} \approx -\frac{N_d^2 }{144} \frac{1}{\epsilon}.\nonumber \\
\end{eqnarray}
For the diagram (f), after integrating over $p_y$ and $q_y$, we find two terms. The first one is
\begin{eqnarray}
(f1)=\int \frac{d^d\mathbf{p}d^d\mathbf{q}}{8 (2\pi)^d} \frac{1}{\mathbf{p}^2+\frac{1}{2}m^2} \frac{1}{\mathbf{p}^2+\mathbf{q}^2+ (\mathbf{p}\!+\!\mathbf{q})^2+\frac{3}{2}m^2}, \nonumber \\
\end{eqnarray}
and the second one is
\begin{eqnarray}
(f2)\!=\!\!\int \frac{d^d\mathbf{p}d^d\mathbf{q}}{8(2\pi)^d} \frac{1}{\left( \mathbf{p}^2+\mathbf{q}^2+ (\mathbf{p}+\mathbf{q})^2+\frac{3}{2}m^2 \right)^2}.\qquad\qquad
\end{eqnarray}
In Schwinger's proper time representation, rewriting the propagator as an integral, we get
\begin{equation}
(f2)=\frac{1}{32} N_d^2\Gamma\left(\frac{d}{2}\right)^2\int_0^\infty dt 
\frac{t}{(3t^2)^{d/2}} e^{-t\frac{3}{2}m^2}\approx \frac{1}{96} N_d^2 \frac{1}{\epsilon}.
\end{equation}
Replacing first $s\to s/3$ and then $t\to t(1-u)$ and $s\to tu$, we have
\begin{eqnarray}
(f1)&=&\frac{1}{8}\frac{1}{(4\pi)^d}\int_0^\infty \! e^{-\frac{3}{2}m^2 s- \frac{1}{2}m^2 t} \frac{dt ds}{\left(3s^2+2st\right)^{d/2}} \nonumber \\
&=&\frac{N_d^2}{32} 3^{d/2-1} \left( \frac{1}{2}m^2\right)^{2-d} \Gamma\left(\frac{d}{2}\right)^{\!\!2} \Gamma(2-d)\nonumber \\ 
&&\times\int_0^1\! \frac{du}{\left(2u-u^2\right)^{d/2}}.
\end{eqnarray}
Calculating the integral with respect to $u$, finally gives
\begin{equation}
(f1)\approx \frac{1}{32}N_d^2\frac{1}{\epsilon^2}+
\frac{1}{64} \left(\ln\frac{4}{3}\right) N_d^2 \frac{1}{\epsilon},
\end{equation}
and all together
\begin{equation}
(f)\approx \frac{1}{32}N_d^2\frac{1}{\epsilon^2}+
\frac{1}{64} \left(\ln\frac{4}{3}+\frac{2}{3}\right) N_d^2 \frac{1}{\epsilon}.
\end{equation}
We also calculate the one of the three-loop diagrams $(h)$, representing it in terms of two parts:
\begin{widetext}
\begin{equation}
(h1)=\frac{1}{16}\int\frac{d^d\mathbf{p}}{(2\pi)^d} \frac{d^d\mathbf{q}}{(2\pi)^d} \frac{d^d\mathbf{k}}{(2\pi)^d} 
\frac{1}{\left((\mathbf{p}+\mathbf{s})^2 +\mathbf{q}^2+(\mathbf{p}+\mathbf{q})^2 +\frac{3}{2}m^2\right) \left((\mathbf{p}+\mathbf{s})^2+\mathbf{k}^2+(\mathbf{p}+\mathbf{k})^2 +\frac{3}{2}m^2\right)}
\end{equation}
and
\begin{equation}
(h2)=\frac{1}{8}\int\frac{d^d\mathbf{p}}{(2\pi)^d} \frac{d^d\mathbf{q}}{(2\pi)^d} \frac{d^d\mathbf{k}}{(2\pi)^d} 
\frac{1}{\left((\mathbf{p}+\mathbf{s})^2 +\mathbf{q}^2+(\mathbf{p}+\mathbf{q})^2 +\frac{3}{2}m^2\right) \left(\mathbf{q}^2+(\mathbf{p}+\mathbf{q})^2 + \mathbf{k}^2+(\mathbf{p}+\mathbf{k})^2+2m^2 \right)},
\end{equation}
where $\mathbf{s}$ is the transferred momentum. The two propagators we rewrite in the Schwinger's proper time representation. Then, transforming the variables to $t_1=tu$, $t_2=t(1-u)$, with the Jacobian: $tdtdu$, and using the two vectors 
$B=(1/3,0,0)$ and $(\mathbf{p},\mathbf{q},\mathbf{k})$ and, finally, the matrix
\end{widetext}
\begin{equation}
A=\frac{1}{3} \left(\begin{array}{ccc} 2 & u & 1-u \\ u & 2u & 0 \\ 1-u & 0 & 2-2u \end{array}\right),
\end{equation}
we obtain
\begin{eqnarray}
(h1)&=&\frac{1}{16}\frac{1}{8}\frac{1}{9} N_d^3\Gamma\left(\frac{d}{2}\right)^3 \int_0^\infty dt \frac{t}{t^{3d/2}} \qquad\qquad\qquad\qquad\,\,\,\nonumber \\
&&\times\int_0^1 du \frac{1}{\left(\textrm{det} A\right)^{d/2}} e^{ -\frac{1}{3}t\mathbf{s}^2 + t BA^{-1}B \mathbf{s}^2-\frac{1}{2} t m^2}.
\end{eqnarray}
Expanding it to the second order in the transferred momentum $\mathbf{s}^2$ results in
\begin{eqnarray}
(h1)&=&-\frac{1}{1152}\frac{1}{9} N_d^3 \Gamma\left(\frac{d}{2}\right)^3 \Gamma\left(3-\frac{3d}{2}\right) \left(\frac{1}{2}m^2\right)^{3d/2-3}
\nonumber \\
&&\times\int_0^1 du \frac{1}{\left(\textrm{det} A\right)^{d/2}}.
\end{eqnarray}
Evaluating this integral gives
\begin{equation}
(h1)=-\frac{1}{864} N_d^3 \frac{1}{\epsilon^2} +\frac{\ln(9/2)}{1728} N_d^3 \frac{1}{\epsilon}.
\end{equation}
In a similar way, we calculate $(h2)$. Putting the two together gives the answer:
\begin{eqnarray}
(h)\approx -\frac{1}{432}N^3_d g^3\frac{1}{\epsilon^2} -\frac{1}{864}N^3_d g^3 
\left(2-\ln\frac{27}{16}\right) \frac{1}{\epsilon}.\quad\quad\quad
\end{eqnarray}
Collecting the diagrams, $-3/2(d)+3/4(e)+3(f)$, and equating it to the counter term,
we find the multiplicative renormalization factor in the second order of the perturbation series:
\begin{eqnarray}
Z_g(g)\!=\!1+\frac{3 N_d g}{8}\frac{1}{\epsilon}+\frac{9 N^2_d g^2}{64}\frac{1}{\epsilon^2}- \frac{3 N^2_d g^2}{64}\!\left(\ln\frac{4}{3} +\frac{2}{3}\right) \frac{1}{\epsilon}.\nonumber \\
\end{eqnarray}
In a similar way, for the $Q^2$ insertion, we find from the sum of the three diagrams:
$-1/2(d)+1/2(e)+1/2(f)$, the multiplicative renormalization factor in the second order 
of the perturbation series:
\begin{eqnarray}
Z_2(g)=1+\frac{N_d g}{8}\frac{1}{\epsilon}+\frac{N^2_d g^2}{32}\frac{1}{\epsilon^2}- \frac{N^2_d g^2}{128}\left(\ln\frac{4}{3} +\frac{2}{3}\right) \frac{1}{\epsilon}.\nonumber \\
\end{eqnarray}
The diagram $-1/6(c)+1/4(h)$ gives immediately the pitch-field renormalization constant in the third order of the perturbation series:
\begin{eqnarray}
Z(g)=1-\frac{N^2_d g^2}{864}\frac{1}{\epsilon}-\frac{N^3_d g^3}{3456}\frac{1}{\epsilon^2}
+\frac{N^3_d g^3}{3456}\left(2-\ln\frac{27}{16}\right) \frac{1}{\epsilon}.\nonumber \\
\end{eqnarray}
Next, using Eq.~(\ref{beta}), we find the $\beta$ function in the second order of the perturbation series:
\begin{eqnarray}
\beta(g)=-\epsilon g +{\textstyle\frac{3}{8}} N_d g^2 -{\textstyle\frac{3}{32}}\left(\ln{\textstyle\frac{4}{3}} +{\textstyle\frac{17}{27}} \right) N_d^2 g^3.\qquad\quad\\\nonumber
\end{eqnarray}
Using Eq.~(\ref{etaG}), we find in the third order of the perturbation series:
\begin{eqnarray}\label{etag}
\eta(g)={\textstyle\frac{1}{432}} N_d^2 g^2 + {\textstyle\frac{1}{1152}} \left(\ln{\textstyle\frac{27}{16}}-2\right) N_d^3 g^3.\qquad
\end{eqnarray}
And using Eq.~(\ref{eta2G}), we find in the second order of the perturbation series:
\begin{eqnarray}\label{eta2g}
\eta_2(g)= -{\textstyle\frac{1}{8}} N_d g +{\textstyle\frac{1}{64}} \left(\ln{\textstyle\frac{4}{3}} +{\textstyle\frac{14}{27}}\right) N_d^2 g^2.\quad
\end{eqnarray}
In the critical point, the free energy settles in into a fixed-point free energy. Solving the fixed-point equation: $\beta(g^*)=0$, for the fixed point vertex $g^*$:
\begin{equation}
N_d g^*={\textstyle\frac{8}{3}}\epsilon+{\textstyle\frac{16}{9}}\left(\ln{\textstyle\frac{4}{3}} +{\textstyle\frac{17}{27}}\right)\epsilon^2,
\end{equation}
we find the critical exponents $\eta$ up to the $\epsilon^3$ and $\eta_2$ up to the $\epsilon^2$ terms of the $\epsilon$ expansion:
\begin{equation}
\eta={\textstyle\frac{4}{243}} \epsilon^2 - {\textstyle\frac{4}{19683}}\left(94+108\ln 2 -135\ln 3\right) \epsilon^3
\end{equation}
and
\begin{eqnarray}
\eta_2=-{\textstyle\frac{1}{3}} \epsilon - {\textstyle\frac{1}{9}}\left(\ln{\textstyle\frac{4}{3}} +{\textstyle\frac{20}{27}}\right) \epsilon^2.
\end{eqnarray}

Let us consider a rescaled Ginsburg-Landau free energy density at a renormalization group fixed point valid for large spin-blocks of size $L$:
\begin{eqnarray}
\frac{1}{2}\tau L^{\eta_2+\eta}\ Q^2 +\frac{g^*}{4!}L^{3\eta/2+d-2}\ Q^4 +\frac{1}{2} L^\eta \left(\nabla Q\right)^2 -hQ.\nonumber \\
\end{eqnarray}
Comparing pairwise two terms in it on the correlation length scale and using the definition of the critical exponent $\nu$ for the correlation length: $\xi\sim 1/|\tau|^\nu$, we obtain
\begin{eqnarray}
(2+\eta_2)\nu &=&1, \nonumber\\ 
\gamma &=& 1-(\eta_2+\eta)\nu, \nonumber\\ 
2\beta &=& 1-\left(2-d +\eta_2 -{\textstyle \frac{1}{2}} \eta \right)\nu, \\ 
\delta &=& 3+\frac{\nu}{\beta} \left(2-d -{\textstyle \frac{3}{2}} \eta \right),  \nonumber\\ 
\alpha &=&-2\beta+1+(\eta_2+\eta)\nu, \nonumber
\end{eqnarray}
where for instance the first line follows from the first and the third terms. We observe that the scaling relationships, specified in Eq.~(\ref{scaling}) in the main text, hold exactly for any value of $\eta_2$. Remembering that $d=2-\epsilon$, we can find the chiral critical exponents in terms of the $\epsilon$ expansion, presented in Eq.~(\ref{critexp}). Although in the first $\epsilon$ order our critical exponents coincide with that of the 3D Ising model, in the next $\epsilon^2$ order they transform into the critical exponents of the 2D Ising model with strong dipolar interactions~[\onlinecite{BrezinZinnJustin}].

The Garel and Doniach model studied in this paper is anisotropic and consequently the correlation length critical exponents are anisotropic, $\nu_x\neq \nu_y$. As can be clearly seen in this Appendix, the critical exponent of the correlation length in $x$-direction, $\nu=\nu_x$, was found. Besides, $\nu_y$ needs not be recalculated, as it is connected to $\nu_x$ through the critical exponent of the correlation function of the helical state, $\eta$. To wit, a complete set of critical exponents is presented in the current manuscript.

\end{document}